\documentclass[a4paper,UKenglish,cleveref, autoref, thm-restate]{lipics-v2021}
\hideLIPIcs  %uncomment to remove references to LIPIcs series (logo, DOI, ...), e.g. when preparing a pre-final version to be uploaded to arXiv or another public repository

\usepackage[frozencache,cachedir=.]{minted}
\usepackage{booktabs}
\usepackage{mathtools}
\usepackage[table,xcdraw,dvipsnames]{xcolor}
\usepackage{amsmath}
\usepackage{amssymb}
\usepackage{graphicx,calc}
\usepackage{xspace}
\usepackage{multicol}
\usepackage{tikz}
    \usetikzlibrary{arrows}

\usepackage[ruled, linesnumbered, noend]{algorithm2e}

\SetCommentSty{mycommfont}
\SetKwRepeat{Do}{do}{while}

\renewcommand{\texttt}[1]{%
  \ifmmode
  \mathtt{#1}%
  \else
  $\mathtt{#1}$%
  \fi
}

\renewcommand{\textsf}[1]{%
  \ifmmode
  \mathsf{#1}%
  \else
  $\mathsf{#1}$%
  \fi
}

\newcommand{\states}{\mathsf{CValue.state}}
\newcommand{\values}{\mathsf{CValue.val}}

\newcommand{\exprs}{\mathsf{Expr}}
\newcommand{\addresses}{\mathsf{Address}}

\newcommand{\contexts}{\mathsf{Context}}
\newcommand{\functions}{\mathsf{Function}}
\newcommand{\stmts}{\mathsf{Stmt}}
\newcommand{\locksets}{\mathsf{Lockset}}

\newcommand{\tool}[1]{\textsc{#1}\xspace}
\newcommand{\racerD}{\tool{RacerD}}
\newcommand{\racerF}{\tool{RacerF}}
\newcommand{\racerX}{\tool{RacerX}}

\newcommand{\racerFsv}{\textsc{RacerF}$_\mathrm{sv-comp}$\xspace}
\newcommand{\racerFunder}{\textsc{RacerF}$_\mathrm{under}$\xspace}
\newcommand{\racerFover}{\textsc{RacerF}$_\mathrm{over}$\xspace}

\newcommand{\goblint}{\tool{Goblint}}

\newcommand{\otwo}{\tool{O2}}

\newcommand{\deagle}{\tool{Deagle}}

\newcommand{\RO}{\textsf{ReadOnly}}
\newcommand{\EX}{\textsf{Exclusive}}
\newcommand{\SH}{\textsf{Shared}}
\newcommand{\SM}{\textsf{SharedMod}}

\title{RacerF: Lightweight Static Data Race\\ Detection for C Code} %TODO Please add

\titlerunning{RacerF: Lightweight Static Data Race Detection for C Code} %TODO optional, please use if title is longer than one line

\author
    {Tom\'{a}\v{s} Dac\'{i}k}
    {Faculty of Information Technology, Brno University of Technology, Czech Republic \and \url{https://www.fit.vut.cz/person/idacik/.en} }{idacik@fit.vut.cz}{https://orcid.org/0000-0003-4083-8943}
    {}

\author
    {Tom\'{a}\v{s} Vojnar}
    {Faculty of Informatics, Masaryk University,
    Brno, Czech Republic
    \and Faculty of Information Technology, Brno University of Technology, Czech Republic \and \url{https://www.muni.cz/en/people/134390-tomas-vojnar} }{vojnar@fi.muni.cz}{https://orcid.org/0000-0002-2746-8792}{}

\authorrunning{Tom\'{a}\v{s} Dac\'{i}k and Tom\'{a}\v{s} Vojnar} %TODO mandatory. First: Use abbreviated first/middle names. Second (only in severe cases): Use first author plus 'et al.'

\Copyright{Tom\'{a}\v{s} Dac\'{i}k and Tom\'{a}\v{s} Vojnar} %TODO mandatory, please use full first names. LIPIcs license is "CC-BY";  http://creativecommons.org/licenses/by/3.0/


 %TODO mandatory; please add comma-separated list of keywords

\relatedversion{} %optional, e.g. full version hosted on arXiv, HAL, or other respository/website
\funding{The research was supported by the Czech Science Foundation project
23-06506S and the FIT BUT internal project FIT-S-23-8151. Tom\'{a}\v{s}
Dac\'{\i}k was also supported by the Brno PhD Talent Scholarship of the Brno
City Municipality. The research involved discussions with Julien Signoles from
the Frama-C team at CEA that were supported by the project VASSAL:
``Verification and Analysis for Safety and Security of Applications in Life''
funded by the European Union under Horizon Europe WIDERA Coordination and
Support Action/Grant Agreement No.
101160022.$\;$\raisebox{0mm}{\includegraphics[width=0.028\textwidth]{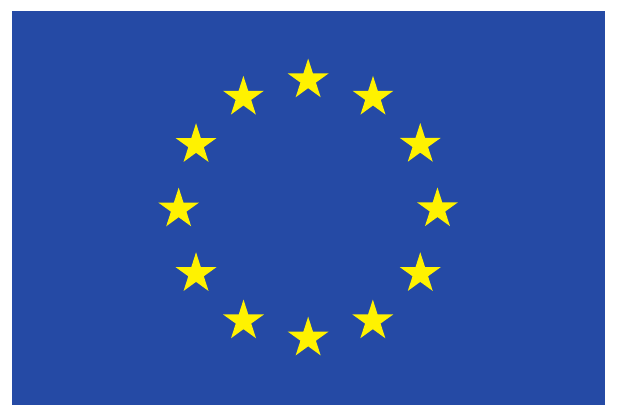}}}

\acknowledgements{
We appreciate discussions on the subject with Julien Signoles
from the Frama-C team at CEA, France and with Adam Rogalewicz from FIT BUT,
Czechia.
}%optional

\nolinenumbers %uncomment to disable line numbering

\EventEditors{Jonathan Aldrich and Alexandra Silva}
\EventNoEds{2}
\EventLongTitle{39th European Conference on Object-Oriented Programming (ECOOP 2025)}
\EventShortTitle{ECOOP 2025}
\EventAcronym{ECOOP}
\EventYear{2025}
\EventDate{June 30--July 2, 2025}
\EventLocation{Bergen, Norway}
\EventLogo{}
\SeriesVolume{333}
\ArticleNo{41}
\begin{document}

\maketitle

\begin{abstract} 
We present \racerF, a novel static analyser for
thread-modular data race detection. The approach behind \racerF exploits static analysis of
sequential program behaviour whose results are generalised for multi-threaded
programs using a combination of lightweight under- and over-approximating
methods. The tool is implemented as a plugin of the Frama-C platform and can leverage several analysis backends,
most notably the Frama-C's abstract interpreter EVA. Although our methods are
mostly heuristic without providing formal guarantees, our experimental
evaluation shows that even for intricate programs, \racerF can provide very
precise results competitive with more heavyweight approaches while being faster
than them.
\end{abstract}

%%%%%%%%%%%%%%%%%%%%%%%%%%%%%%%%%%%%%%%%%%%%%%%%%%%%%%%%%%%%%%%%%%%%%%%%%%%%%%%%
\section{Introduction}
%%%%%%%%%%%%%%%%%%%%%%%%%%%%%%%%%%%%%%%%%%%%%%%%%%%%%%%%%%%%%%%%%%%%%%%%%%%%%%%%

Data races in concurrent programs belong among the most nasty bugs in computer
programs: they are easy to cause, hard to spot during code inspection, and hard
to catch by testing as they can manifest only very rarely.
Due to this, a lot of effort has gone into designing as safe as possible
libraries and languages for writing concurrent programs, but, on one hand, many
programs are written using languages without such safety features (with the C
language being a prominent example), and on the other hand, even in the
safety-oriented languages such as Rust, unsafe programming is usually allowed
and sometimes used, e.g., for performance reasons.

A lot of work has also been invested in designing various static as well as
dynamic methods of finding data races and/or proving their absence.
However, the complex behaviour of concurrent programs reflects in that such
methods do often have problems with scalability or precision.
In this paper, we present a novel tool \racerF for thread-modular data race
detection with the aim of providing a suitable balance between the scalability
and precision of the analysis.
We then also present our experience from running the tool and comparing it with
several other data race detectors.

%In this paper, we contribute to the ongoing search for a suitable balance
%between the scalability and precision of data race detection by proposing a
%novel static analysis for thread-modular data race detection.

The approach implemented in \racerF relies on leveraging results of static
analysis of the sequential behaviour of program threads for finding data races
in multi-threaded programs.
We use two strategies of combining information resulting from the analysis of
the sequential behaviour of program threads: one under-approximates and one
over-approximates the concurrent behaviour of the threads running in parallel.
We combine these strategies  to obtain a more precise analysis such that the
under-approximation is used to find data races while the over-approximation is
used to show their absence.

%We have implemented our approach in a new data race detector for C programs with
%POSIX threads (pthreads).
%%
%The tool is called RacerF and it is implemented as a plugin of the Frama-C
%platform \cite{Frama-C}.

\racerF is implemented as a plugin of the Frama-C platform \cite{Frama-C1, Frama-C2}
and is in particular designed for finding data races in C programs with POSIX
threads (pthreads).
\racerF can use several analysis backends implementing the underlying analysis
of the sequential behaviour of program threads and differing in their precision
and scalability.
One of them is built on top of the Frama-C's value analysis plugin EVA based on
abstract interpretation \cite{EVA}, one is purely syntactic, and the last one is
based on \textsc{Alias}, Frama-C's plugin for may-alias and points-to analysis \cite{alias}.

We have performed experiments with \racerF on a number of benchmarks coming
both from the SV-COMP competition in software verification, which includes both
synthetic benchmarks as well as benchmarks derived from real-world code, as well
as on a number of student projects in an advanced course of operating systems.
Although the approach behind \racerF is mostly heuristic without providing
formal guarantees, our experimental evaluation shows that even for intricate
programs, \racerF can provide very precise results competitive with more
heavy-weight approaches while being faster than them.

%%%%%%%%%%%%%%%%%%%%%%%%%%%%%%%%%%%%%%%%%%%%%%%%%%%%%%%%%%%%%%%%%%%%%%%%%%%%%%%%
\subsection{Approach Overview}
%%%%%%%%%%%%%%%%%%%%%%%%%%%%%%%%%%%%%%%%%%%%%%%%%%%%%%%%%%%%%%%%%%%%%%%%%%%%%%%%

\begin{figure}[t!]
    \centering
    \includegraphics[width=\linewidth]{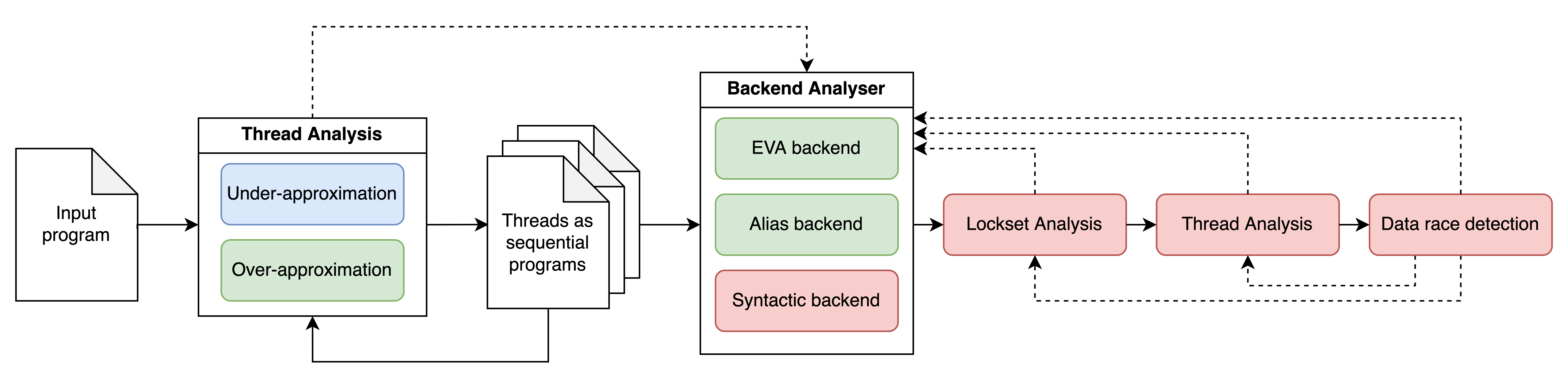}
    \caption{An overview of the workflow of \racerF. The phases with under- and over-approximation guarantees are marked by blue and green colour, respectively. The phases designed without guarantees are marked by red colour. The solid lines represent the workflow of individual analysis phases. The dashed lines represent queries on the results of other phases.}
    \label{fig:analysis_pipeline}
\end{figure}

An overview of the workflow of \racerF is illustrated in Fig. \ref{fig:analysis_pipeline}.
The core component is a \emph{backend analyser} which answers several kinds of
queries based on analysing individual threads as sequential programs (the
general interface and several different backend analysers are
described in Section \ref{sec:backend}).
The \emph{thread analysis} (Section \ref{sec:thread-analysis}) incrementally
discovers new program threads and analyses them using the chosen backend analyser.
Since the discovery of a new thread may add additional information, the process
of thread discovery and their analysis using the backend analyser is
repeated until a fixpoint is reached.
Then, two helper analyses are run to deal with checking whether two memory
accesses may happen in parallel  -- a \emph{lockset analysis} checks which
accesses are guarded by which locks and an \emph{active-threads analysis} checks
when threads are created and joined (Section
\ref{sec:lockset_and_thread_analyses}). Those analyses query results of the
backend analysis, mainly for possible values of parameters of locking and
threading functions.
Finally, \emph{data race detection} (Section \ref{sec:race-detection}) is
performed. This phase checks memory accesses (using information provided by
the backend analyser) and tracks whether they need to be protected and, if
so, whether they indeed are. The detected issues are then classified  into may-
and must-data-races. Must-data-races are reported to the user, while an absence
of may-data-races is used to claim program race-free.

Fig. \ref{fig:analysis_pipeline} also depicts the approximation guarantees
provided by the individual phases. Note that the combination of under- and over-approximating methods may seem problematic from the point of view of what can be guaranteed to hold for the produced results. However, we do not aim at providing guarantees in the sense of formal verification. Our aim is to achieve high efficiency and precision in practice even though some false alarms may be missed or some bugs not identified.

%%%%%%%%%%%%%%%%%%%%%%%%%%%%%%%%%%%%%%%%%%%%%%%%%%%%%%%%%%%%%%%%%%%%%%%%%%%%%%%%
\section{Related Work}
%%%%%%%%%%%%%%%%%%%%%%%%%%%%%%%%%%%%%%%%%%%%%%%%%%%%%%%%%%%%%%%%%%%%%%%%%%%%%%%%

%\subparagraph*{Static analysis.} 
\racerX \cite{racerx} is a static analyser for C based primarily on a lockset
analysis. The tool has been successfully used to find bugs in the kernels of
several operating systems. Compared to our approach, \racerX uses a very simple
pointer analysis for the representation of locks only. Further, \racerX uses many
heuristics in individual analysis phases to deal with analysis mistakes while
\racerF tries to achieve precision using careful ranking. \racerX is
unfortunately not available for an experimental comparison.
\racerD~\cite{racerd} is a compositional analyser implemented in the Infer
framework. It provides scalability by analysing each function only once. Unlike
us, \racerD targets Java programs with structural locking, assuming that all
locking operations are balanced. Both \racerX and \racerD focus on
under-approximation only. \otwo \cite{o2} is another static checker building on
the notion of origins unifying threads and events. The approach was implemented
for C, C++, and Java/Android; and used to find many previously unknown bugs in
real-world code. However, as we show in our experiments, it can still miss many
data races in smaller but real-world programs.

On the other side, there are tools focused on soundness, rather than data race
detection. Goblint \cite{goblint} and Astrée~\cite{astree2,astree1} rely on a
thread-modular abstract interpretation to soundly over-approximate thread
behaviour. The work \cite{goblint} further proposes a way to handle common
locking schemes in Linux driver devices. In our approach, we are not interested
in the analysis of multithreaded programs in general and we can consequently
afford more drastic approximations that seem to be sufficient for data race
detection.  We also focus on under-approximation to detect bugs.

As an alternative to static data race detection, various approaches based on
\emph{testing} and \emph{dynamic analysis} can be used. A big challenge for
these approaches is the fact that errors like data races can manifest in very
rare behaviours only, which are difficult to catch, even through many repeated
test runs. Coverage of such rare behaviours can be improved using approaches
such as \emph{systematic testing}~\cite{schedspec12} or \emph{noise-based
testing}~\cite{contestframework03,noise15}. Another way is to use
\emph{extrapolating dynamic checkers}, such
as~\cite{Eraser,goldilocks07,fasttrack09}, which can report warnings about
possible errors even if those are not seen in the testing runs, based on
spotting their symptoms. However, even though such checkers have proven quite
useful in practice, they can, of course, still miss rarely occurring errors.
Moreover, monitoring a run of large software through such checkers may be quite
expensive too.

% Combining static and dynamic analysis may then leverage advantages of both
% approaches, allowing static analysis to be more focused and hence more
% efficient~\cite{noise15,bigfoot17}.

Approaches based on \emph{model checking}, i.e., exhaustive state-space
exploration, can guarantee the discovery of all potentially present errors --
either in general or at least up to some bound, often given in the number of
context switches. However, so far, the scalability of these techniques is not
sufficient to handle large industrial code, even when combined with methods such
as \emph{sequentialisation}~\cite{lal-reps-08,lazy-seq-16} or advanced techniques
based on \emph{automata} \cite{automizer, gemcutter} or \emph{SMT encodings}
%\tv{Tady by to chtelo alespon nejaky privlastek k tomu SMT encodings. Careful? A
%SMT encodings ceho vlastne?} 
\cite{dartagnan, deagle, zord}.
%and others used in the newest tools such as Deagle~\cite{deagle}.
Indeed, this shows up in our experiments too.

%%%%%%%%%%%%%%%%%%%%%%%%%%%%%%%%%%%%%%%%%%%%%%%%%%%%%%%%%%%%%%%%%%%%%%%%%%%%%%%%
\section{Preliminaries}
%%%%%%%%%%%%%%%%%%%%%%%%%%%%%%%%%%%%%%%%%%%%%%%%%%%%%%%%%%%%%%%%%%%%%%%%%%%%%%%%

We now introduce several basic notions that we use when describing the analysed programs and the analyses proposed.

%-------------------------------------------------------------------------------
\subparagraph*{Concurrency model.}
%-------------------------------------------------------------------------------

We primarily target C programs using threads and mutexes in the style of the
POSIX threads (pthreads) execution model. We use the term \textit{thread} to
mean thread's entry point, i.e., an identifier of the function from which the
given thread is started. In the examples, we shorten the calls of
locking, unlocking, thread-creating, and thread-joining functions as
\texttt{lock(...)}, \texttt{unlock}(...), \texttt{create}(...), and
\texttt{join}(...), respectively. We further implicitly assume that all
locks are properly initialised and omit the arguments that are not relevant,
e.g., we write just \texttt{create(t1)}, where \texttt{t1} is an entry point,
when the thread id, attributes, and argument are not relevant for the example.

%-------------------------------------------------------------------------------
\subparagraph*{The CValue domain and memory addresses.}
%-------------------------------------------------------------------------------

Throughout the paper, we frequently refer to the \textsf{CValue} abstract domain
\cite{EVA,Frama-C1} implemented in Frama-C and used by the abstract
interpretation of EVA. We are primarily interested in the representation of
memory addresses. A memory address is represented as a pair $(b, o)$ where $b$
is a \textit{base address} (in the following referred to simply as a
\textit{base}) and $o$ is an offset. Each variable (global, local, and also
formal) has its associated base address. Dynamic allocations are modelled by
creating so-called \textit{dynamic bases} which come in two flavours -- a
\textit{strong} dynamic base corresponds to the result of precisely one call to
an allocation function; on the other hand, \textit{weak} memory bases may
represent multiple allocated addresses (typically, they are the consequence of
allocations in loops). Offsets are represented by a dedicated integer domain,
the details of which are not important for this paper. We use
\textsf{CValue.val} for the domain representing possible values of a single
variable and \textsf{CValue.state} for the domain representing program state,
i.e., possible values for multiple variables. We use \textsf{Addresses} to
denote the aforementioned set of memory addresses. In the rest of the paper, the
operations \emph{join} and \emph{widening} as well as \emph{top values} refer to
those on the \textsf{CValue} domain.

%-------------------------------------------------------------------------------
\subparagraph*{Analysis contexts.}
%-------------------------------------------------------------------------------

We call an \emph{analysis context} a triple $(thread, calls, stmt)$ consisting
of a thread, a sequence of the latest $n$ calls (each of them formed by a call
site and the function called), and a program statement. Our data race analysis
relies on various underlying analyses presented below to compute an abstract
representation of thread states reachable by performing the given statement for
each analysis context. Notice that a context holds no information about how the
thread was created nor how its execution was interleaved with other threads. The
empty call sequence is denoted by $\varepsilon$. The number $n$ is a parameter
of the analysis set to 1 by default to handle simple lock/unlock wrapper
functions. We use $\contexts$ to denote the set of all possible contexts.

\begin{figure} \centering
    %%$$ %   \mathsf{type}\;\mathsf{Res} \;\; \text{(Type representing backends'
    %%   result)} %$$
    \begin{align*} \mathsf{analyse\_thread} &: \mathsf{Function} \times
    \mathsf{CValue.state} \xrightarrow{} \mathsf{Result} && \hspace{-5pt}\text{(sequential
    analysis of a thread)}\\ \mathsf{state} &: \mathsf{Result} \times \contexts
    \xrightarrow{} \mathsf{CValue.state} && \hspace{-5pt}\text{(state before a statement)}\\
    \mathsf{value} &: \mathsf{Result} \times \contexts \times
    \mathsf{Expr}\xrightarrow{} \mathsf{CValue.val} && \hspace{-5pt}\text{(non-pointer
    expression values)}\\ \mathsf{value\_ptr} &: \mathsf{Result} \times
    \contexts \times \mathsf{Expr}\xrightarrow{} 2^{\addresses} &&
    \hspace{-5pt}\text{(pointer expression values)}\\ \mathsf{functions} &: \mathsf{Result}
    \times \contexts \times \exprs \xrightarrow{} 2^{\functions}
    &&\hspace{-5pt}\text{(function pointer resolution)}\\ \mathsf{accesses} &:
    \mathsf{Result} \times \contexts \times \mathsf{Expr} \xrightarrow{}
    (2^\mathsf{Address})^2 && \hspace{-5pt}\text{(memory reads and writes)}
    \end{align*}

    \caption{The interface of the backend analysis used by the proposed analysis
    of data races. The function \textsf{analyse\_thread} performs the analysis of a
    thread as a sequential program given by its entry point, starting from the
    given initial state. The other functions answer queries based on the
    results computed by \textsf{analyse\_thread} (specific for different backends).
    Note that the signatures do not include statements as those are already
    included in contexts.}
    \label{fig:backend_sig}
\end{figure}

%%%%%%%%%%%%%%%%%%%%%%%%%%%%%%%%%%%%%%%%%%%%%%%%%%%%%%%%%%%%%%%%%%%%%%%%%%%%%%%%
\section{Backend Analysis} \label{sec:backend}
%%%%%%%%%%%%%%%%%%%%%%%%%%%%%%%%%%%%%%%%%%%%%%%%%%%%%%%%%%%%%%%%%%%%%%%%%%%%%%%%

Our data race analysis relies on a backend analysis of sequential program
behaviour to get answers for several kinds of queries. Those include
\emph{function pointer resolution} (to determine which functions are to be
analysed when and to determine thread entry points passed to thread create
functions), getting \emph{points-to information} (to determine targets of
(un)lock operations, to perform escape analysis, and, most importantly, to
determine on which addresses to look for data races), getting information on
\emph{memory accesses}, and characterizing \emph{values of integer variables}
(to determine whether offsets of two accesses overlap, and to check whether a \textsf{trylock} operation succeeds in the given branch
of code).

To offer a selection of multiple backend analyses differing in their
scalability, precision, and ease of use, we have implemented several such
analyses. The interface a backend needs to implement is shown in Fig.
\ref{fig:backend_sig}. The function \textsf{analyse\_thread} runs the analysis of a
thread as a sequential program starting from the provided state of global
variables and the thread's argument. Based on the results, the different queries
are then answered.

%-------------------------------------------------------------------------------
\subparagraph*{Evolved Value Analysis (EVA) backend.}
%-------------------------------------------------------------------------------

This backend is based on EVA \cite{EVA}, a value analysis plugin of Frama-C
based on abstract interpretation. EVA provides a combination of multiple abstract domains and an API to answer all the needed queries, and our backend
is just a thin wrapper over EVA, which implements caching, simplification, and
answering some trivial queries on its own. The interface in Fig.
\ref{fig:backend_sig} is in fact inspired by the API of EVA, and the other
backends mimic it. In particular, the EVA's API implements queries for call
stacks which can be easily obtained from our contexts.

%We set the context depth
%used by our data race analysis as the call stack depth for EVA so that it does
%not compute more precise information than is needed.

%There is, however, one catch -- EVA supports single-threaded programs only. In
%Section \ref{sec:thread-analysis}, we give more details on how we use EVA to
%compute value analysis in a multi-threaded setting.

%-------------------------------------------------------------------------------
\subparagraph*{Syntactic backend.}
%-------------------------------------------------------------------------------

This backend provides a lightweight analysis based only on syntactical
information in the style of \racerX\cite{racerx} or \racerD\cite{racerd}. The
implementation of \textsf{analyse\_thread} is simply an empty function and the functions
\textsf{state} and \textsf{value} always return the top value. The function
\textsf{value\_ptr} searches the expression and extracts bases of the expected
type. This is inspired by the insight that in some programs only global locks
are accessed using $\mathtt{lock(\&m)}$ and $\mathtt{unlock(\&m)}$
(where \texttt{m} is a global variable), and it is
thus sufficient to simply extract the base of $\mathtt{m}$. The function
\textsf{memory\_accesses} is implemented similarly with a more detailed analysis
of an expression to determine whether a base is being read or written to it.
Finally, function pointers are over-approximated by taking each function
assigned in the program. In the case of thread entry points, this can be further
refined by taking only functions matching the thread signature (i.e.,
\mbox{\texttt{void *(*fn) (void *arg)}} for \textsf{pthreads}).

%-------------------------------------------------------------------------------
\subparagraph*{Alias backend.}
%-------------------------------------------------------------------------------

The aim of the third backend is to provide a middle ground between the precision
of the EVA and the scalability of the syntactic backend. It relies on another
\mbox{Frama-C} plugin called Alias \cite{alias} which performs a may-alias
analysis using a variant of the classical Steensgaard’s algorithm
\cite{Steensgaard}. For queries not related to pointer expressions
(\textsf{state} and \textsf{value}), the top values are returned. For the other
queries, since Alias does not have an API to answer our queries directly, we take
the result of the syntactic backend and replace each base by a canonical
representative of its alias equivalence class. In particular, for memory
accesses, we ensure that a selected canonical address for an alias class is
vulnerable to data races if such an address exists (e.g. because its base
corresponds to a global variable or a local variable that escaped; see Section
\ref{sec:race-detection} for more details). Alias also does not allow starting
with a user-supplied set of aliases (as expects the signature of our
\textsf{analyse\_thread} function). We therefore need to perform an additional
saturation of the aliases set for all queries, as demonstrated by the example in the
next paragraph. The Alias plugin is still relatively new (it lacks support,
e.g., for the dynamic memory allocation) and thus our backend over it is
also experimental.

\begin{figure}
\begin{multicols}{2}
  \begin{minted}[numbersep=-6pt, obeytabs=true,tabsize=1,linenos]{c}
   void *t1(void *arg1) {
     int *x = (int *)arg1;
     (*x)++;
   }

   void *t2(void *arg2) {
     int *y = (int *)arg2;
     (*y)++;
   }
   int main() {
     int data = 0;

     create(t1, &data);
     create(t2, &data);
   }
    \end{minted}
    \phantom{a}\\
    \phantom{a}
\end{multicols}

  \caption{An example of a program for which the Syntactic backend fails to find
  a data race.}
  \label{fig:backend_example}
\end{figure}

\begin{example}
The advantage of the Alias backend over the syntactic backend is demonstrated in
Fig. \ref{fig:backend_example}. The program contains a data race on an address
of the variable \texttt{data} that escapes its scope in the \texttt{main}
function by passing it as an argument to the threads \texttt{t1} and
\texttt{t2}. Both threads share the same code, which creates a local alias for
this address. This is a very common pattern when thread arguments are used --
since the type of the argument is generic \texttt{void*}, it needs to be casted
to its actual type. The address is then written to by both threads using their
local aliases. While the syntactic backend would not report any relevant access
for data race detection, the Alias backend would compute aliases for
\texttt{x} and \texttt{y} as $\{\mathtt{x}, \mathtt{arg1}\}$ and $\{\mathtt{y},
\mathtt{arg2}\}$, respectively. Our backend then performs a saturation of both
sets by adding aliases $\{\mathtt{arg1}, \mathtt{\&data}\}$ and
$\{\mathtt{arg2}, \mathtt{\&data}\}$, respectively. Finally, \texttt{\&data} will be
selected as a canonical base, because it is vulnerable to a data race.
\end{example}

%%%%%%%%%%%%%%%%%%%%%%%%%%%%%%%%%%%%%%%%%%%%%%%%%%%%%%%%%%%%%%%%%%%%%%%%%%%%%%%%
\section{Thread Analysis}
\label{sec:thread-analysis}
\enlargethispage{8mm}
%%%%%%%%%%%%%%%%%%%%%%%%%%%%%%%%%%%%%%%%%%%%%%%%%%%%%%%%%%%%%%%%%%%%%%%%%%%%%%%%

In this section, we propose our approach on how to leverage an analyser for
sequential programs to analyse programs with multiple threads. The design of our
approach is based on the observation that answers to our queries described in
Section \ref{sec:backend} on objects relevant for our analysis are often not too
much affected by thread interleavings. We design two strategies for combining
results of the analyses of the sequential behaviour of individual threads that
under- and over-approximate the overall behaviour of the concurrent program
consisting of the threads (not including behaviour possibly caused by errors in the code, i.e., for example undefined behaviour after a data race).
%provided that the behaviour of the program is not restricted by some kind of a concurrency error such as a deadlock\footnote{For a program with a deadlock, our
%under-approximation would over-approximate the reachability by ignoring a
%possibility of the deadlock} or even data race\footnote{For a program with a data race, our over-approximation would ignore consequences of possible undefined behaviour. We note that \racerF can detect data races even without observing their effects during thread analysis.}, which the backend analyser of sequential behaviour does not take into account.) 
Furthermore, these strategies can subsequently be
combined to increase the precision of the resulting analysis by using
under-approximation to find races while using over-approximation to show their
absence. However, as we show in our experimental evaluation in Section
\ref{sec:experiments}, both of our strategies are often sufficiently precise on
their own.

The core idea of our thread analysis is similar to that of the Frama-C's
proprietary Mthread plugin described in \cite{mthread}. Starting with the
\texttt{main} thread, we iteratively build a thread-creation graph that captures
the parent-child relationship among the so-far discovered threads. In each
iteration, we add newly discovered threads
and perform the analysis (potentially multiple times) of each
thread as a sequential program (using the \texttt{analyse\_thread} function of the
selected backend). This is done by solving a system of equations over the
initial states of the threads given by
whose construction we describe below. However, before that, we first describe
the graph construction algorithm.

The section is relevant mainly for our EVA backend and partly for our Alias
backend. For the Syntactic backend, only the thread-creation graph construction
is relevant (because no analysis of threads is performed).

\begin{algorithm}[t]
\SetKwInOut{Input}{Input}
\SetKwInOut{Output}{Output}
\SetKwInOut{Data}{Data}
\caption{Thread analysis}
\label{alg:thread-analysis}
\Input{Program entry point $\mathsf{main}$}
\Input{Initial values of global variables $\mathsf{GlobalsVals} \in \mathsf{Cvalue.state}$}
\Data{Initial states of threads $I : T \xrightarrow{} \mathsf{Cvalue.state}$}
\Output{Thread-creation graph $G = (T, \xrightarrow{})$}
\Output{Results of analysis of sequential behaviour of the threads $R : T \xrightarrow{} \mathsf{Result}$}
\DontPrintSemicolon

$i = 0$ \;

$G_0 = (\{\mathsf{main}\}, \emptyset)$ \tcp*{Initial graph containing 
\textsf{main} only}

$I_0 = \{\mathsf{main} \mapsto \mathsf{GlobalsVals}\}$ \tcp*{Initial state of the \textsf{main} thread}

$R_0 = \{\mathsf{main} \mapsto \mathsf{Backend.analyse\_thread}(\mathsf{main}, \mathsf{GlobalsVals}) \}$ \tcp*{Analyse \textsf{main}}

\Do{$G_{i-1} \neq G_i$}{
    $i = i + 1$ \;
    
    $G_i = G_{i-1}$ \;
    
    \tcc{Update the graph with reachable threads.}
    %$G_i = \mathsf{computeThreadGraph}(G_{i-1})$
    \ForEach{\emph{thread represented as a vertex} p \emph{in} $G_{i-1}$}{

        %$Q = \mathsf{Backend.analyse}(parent, I_{i-1}(parent))$\;
        
        \ForEach{\emph{stmt} s \emph{of the form} $\mathsf{thread\_create}(f)$
        \emph{syntactically reachable by} $p$}
        {
            $children = \mathsf{Backend.functions}(R_{i-1}(p), (p, \varepsilon, s), f)$ \;
            $G_i = G_i \cup \{p \xrightarrow{s} c \;|\; c \in children\}$
        }
    }

    \tcc{Build equations wrt. the chosen approx. strategy.}
    $\Phi_i = \mathsf{ConstructEquations}(G_i)$

    \tcc{Solve the system of equations.}
    $I_i = \mathsf{SolveEquations}(\Phi_i, I_{i-1})$\;

    \tcc{Analyse each thread as a sequential program.}
    $R_i = \{t \mapsto \mathsf{Backend.analyse\_thread}(t,I_i(t)) \;|\; t \in \mathsf{dom}(I_i)\}$
    
}
\Return{$G_i, R_i$}
    
\end{algorithm}

%===============================================================================
\subsection{Construction of the Thread-Creation Graph}
%===============================================================================

The method is given in Algorithm \ref{alg:thread-analysis}. The output of the
algorithm is (1) a thread-creation graph $G = (T, \rightarrow)$ where $T$ is the
set of all discovered threads and $\rightarrow$ is a labelled edge relation such
that $p \xrightarrow{s} c$ means that the parent $p$ creates a \mbox{child $c$}
at the program statement~$s$, and (2) a mapping $R : T \xrightarrow{}
\mathsf{Result}$ containing results of the analysis of the behaviour of each
thread.  The algorithm further operates with the mapping $I : T \xrightarrow{}
\states$ representing the initial states of discovered threads\footnote{In the
implementation, the product $\states \times \values$ is used as the domain of
the initial states to represent the values of global variables and the value of
the thread's argument. However, for simplicity of the presentation, we may
assume that the argument is simply included among the global variables.}. Once
we derive a stable mapping $I$ (i.e., a fixpoint of the thread's initial
states), we have also reached a fixpoint of the results $R$, needed to
answer all the queries issued by our analysis.

The algorithm starts with a graph that contains only the \textsf{main} thread
whose initial state is given by the initial values of global variables. With
this initial state, \textsf{main} is analysed on line 4.

In the main loop of the algorithm starting on line 5, the graph is updated by
adding threads created from (syntactically) reachable statements. Notice that
the evaluation of an expression $f$ on line 10 is performed without using any
calling context. This is because our analysis, in particular the system of
equations in Sections \ref{sec:thread-analysis_under} and
\ref{sec:thread-analysis_over}, is not very sensitive to thread-create wrappers.
Consequently, we can construct smaller graphs by distinguishing edges only by
statements and not by calling contexts.

After a new iteration of the graph $G_i$ is constructed, we build a system of
equations, one for each discovered thread, given by $G_i$. Each of these
equations describes how an initial state of a thread can be computed using
initial states of other threads. Then, the system is solved using initial states
from the previous iteration $I_{i-1}$ to obtain new initial states $I_{i}$. The
solution may need an application of widening operators -- we provide more details
on our implementation of the equation solving in Section~\ref{sec:experiments}. Based on
the new initial states, we compute new results $R_i$ on line 14.

This loop is repeated until the thread-creation graph stabilises. Overall, the
algorithm runs two fixpoint computations -- the outer updates the graph with new
threads, while the inner (equation solving on line 13) performs analysis on the current graph, and can
thus help to discover previously unknown threads in new iterations.

Below, we discuss how the aforementioned equations are constructed.

%===============================================================================
\subsection{Under-Approximating Thread Analysis}\label{sec:thread-analysis_under}
%===============================================================================

We use the following under-approximated semantics of thread creation. A thread
is started with the initial state corresponding to the join of initial states of
its create statements, and its writes to global variables are not propagated to
other threads. The creation of threads then roughly corresponds to calling its
entry point in a “sandbox” that makes all of its actions invisible for other
threads. Formally, given a thread-create graph $G_i$ with \mbox{edges
$\xrightarrow{}$}, we compute the initial state of a thread $t$ (other than
$\texttt{main}$ -- the initial state of $\texttt{main}$ is given by the initial
states of global variables) as follows:

\begin{align*} I(t) \;\;\triangleq\;\; \bigsqcup_{p \xrightarrow{s} \;t}
\mathsf{let}\;R = \mathsf{Backend.analyse\_thread}(p, I(p)) \; \mathsf{in}\;
\mathsf{Backend.state}(R, (p, \varepsilon, s)), \end{align*}

\noindent i.e., we take all statements $s$ creating $t$, and for each of them,
we recompute the corresponding parent $p$ and take its state at the statement
$s$, returning the join of all those values. In this way, the creation of a
thread does not affect its parent at all.

The motivation for this strategy is straightforward -- for acyclic
thread-creation graphs, it is enough to analyse each entry point only once
(provided caching is used for the \texttt{analyse\_thread} function). On the other
hand, there are two main sources of possible imprecisions -- thread
interleavings are not taken into account, and parents do not see effects of
their children. The latter can sometimes result in seemingly weird false
negatives when performing data race detection, as demonstrated in Listing
\ref{fig:thread-example-under}. In the program, the \texttt{main} thread first
creates a thread \texttt{t1} which allocates an integer value and then a thread
\texttt{t2} (created twice) accesses this value without any synchronization.
However, we cannot detect this data race using EVA\footnote{We could detect the memory access with the other backends that do not care about memory initialization.} since in our
under-approximated setting, \texttt{t2} will never see the allocation, and in
such a situation, EVA will mark  the memory access on line 6 as surely invalid
and will not report any memory access for it.

\begin{figure}[t]
\begin{subfigure}{0.45\textwidth}
\begin{minted}[numbersep=-6pt, obeytabs=true, linenos=true]{c}
   int *g;
   void *t1() {
     g = malloc(...);
   }
   void *t2() {
     (*g)++;
   }
   int main() {
     create(t1); join(t1);
     create(t2); create(t2);
   }
\end{minted}
\caption{} \label{fig:thread-example-under}
\end{subfigure}
\begin{subfigure}{0.5\textwidth}
\begin{minted}[numbersep=-6pt, obeytabs=true, linenos]{c}
   int g = 1;
   void *thread() {
     lock(L);
     if (g == 1) {
       unlock(L); return NULL;}
     g = 2;
   }
   int main() {
     create(t1);
     lock(L); g = 0; g = 1; unlock(L);
   }
\end{minted}
\caption{} \label{fig:thread-example-over}
\end{subfigure}
    \caption[]{Examples of programs on which (a) the under- and (b) over-approximating thread analysis reports a false
    negative and a false positive, respectively.}
    \label{fig:thread_examples}
\end{figure}

%===============================================================================
\subsection{Over-Approximating Thread Analysis} \label{sec:thread-analysis_over}
%===============================================================================

While for the under-approximating strategy, we assumed that no changes done by
threads are propagated, in our over-approximating strategy, we assume that every
thread can see all possible changes (even those that should be invisible due to some
synchronization method). For this method to work correctly, we also need to
perform a simple transformation of the analysed program, which we discuss,
together with a motivation example, later in this section.

We compute the initial state of a thread~$t$ (here also including the
\texttt{main} thread) as the join of the states our analysis has encountered so
far. Formally, given a thread graph $G_i$ with a set $T$ of currently known threads, the initial values of global variables $V = \mathsf{GlobalsVals}$, and a set of all program statements $\stmts$, we proceed as
follows:

$$I(t) \triangleq V \;\sqcup \hspace*{-5mm}\bigsqcup_{s \in \stmts,\; t' \in T}
\hspace*{-6mm}\mathsf{let}\;R = \mathsf{Backend.analyse\_thread}(t', I(t')) \;
\mathsf{in} \; \mathsf{Backend.state}(R, (t', \varepsilon, s)),$$

\noindent i.e., we take the join of the states of all statements
for each currently discovered thread.

To solve such a system, more than one sequential analysis of each entry point
and an application of widening is needed to converge\footnote{We apply widening
quite aggressively by performing it already after the second iteration to reach
a fixpoint as soon as possible.} (cf.  Section~\ref{sec:experiments}).  Observe
that the equations do not depend on edges of the thread-creation graph, just on
its vertices.

As we also consider context switches that cannot happen due to synchronization
(even those that cannot happen because a child was not yet created, e.g., when
computing the initial state of the main thread and already using states of its
children), we may later report false data races as demonstrated in Listing
\ref{fig:thread-example-over}. In this example, we will compute the initial
state of \texttt{thread} to be $\{\mathtt{g} \mapsto \{0,1,2\}\}$. However,
the value $0$ is never visible to \texttt{thread} because the sequence of
updates on line 10 (as well as the condition on line 4) is performed atomically,
resulting in no change of \texttt{g}. Thus, the memory access on line 6,
which would be flagged as a data race (together with one of the accesses on line
10), is in fact unreachable.

\definecolor{lightyellow}{rgb}{1.0, 1.0, 0.5}

\begin{figure}
\begin{subfigure}{0.45\textwidth}
\begin{minted}[numbersep=-6pt, obeytabs=true,linenos, tabsize=1]{c}
   atomic_int g;

   void *t() {
     g++;
   }

   int main() {
     g = 0;
     create(t);
     if (g == 1)
       ...
   }
\end{minted}
\caption{Original program.} \label{fig:overapprox_fail_orig}
\end{subfigure}
\begin{subfigure}{0.30\textwidth}
\begin{minted}[obeytabs=true, linenos=false, tabsize=1, highlightlines={4,8}, highlightcolor=lightyellow]{c}
atomic_int g;

void *t() {
  g = (*) ? g : g + 1;
}

int main() {
  g = (*) ? g : 0;
  create(t);
  if (g == 1)
    ...
}
\end{minted}
\caption{Transformed program.} \label{fig:oveapprox_fail_transormed}
\end{subfigure}

  \caption{An example of a program for which the over-approximating strategy
  fails to compute a sound over-approximation unless the program is
  transformed (\texttt{*} denotes non-deterministic condition).}

  \label{fig:overapprox_fail_example}
\end{figure}

Although both examples in Fig. \ref{fig:thread_examples} are taken from publicly
available benchmarks, we believe that they are quite rare in practice. Moreover,
our strategies can be combined by allowing the under-approximation to only
report errors and the over-approximation to only report that no error has been
found. Otherwise, an unknown verdict will be reported, which would be also the
case in both examples in Fig.~\ref{fig:thread_examples}.

%-------------------------------------------------------------------------------
\subsubsection{Program transformation}
%-------------------------------------------------------------------------------

The presented over-approximating strategy is not in itself sufficient to compute
a sound over-approximation of multi-threaded behaviour as demonstrated on the
program in Figure~\ref{fig:overapprox_fail_example}. In the example, we are
particularly interested in the block inside the condition on line 10, which may
contain some important event for our analyser. This code is unreachable when the
main thread is analysed as a sequential program regardless of the initial state.
Indeed, our over-approximation strategy would compute the initial state as
\mbox{$\{\texttt{g} \mapsto [0, \mathsf{max\_int}]\}$}, but already on line~8, this
state would be modified to $\{\texttt{g} \mapsto [0, 0]\}$ and the
over-approximation would be lost.

We solve the above problem using a simple source code transformation rather than
modifying the internal state of the sequential analyser (so we do not need to
make further requirements on the interface of our backends). The idea behind our
solution is that it is sufficient to ensure that each program instruction is
\textit{extensive}, i.e., if $S$ and $S'$ are the sets of concrete states  before and
after the instruction, respectively, then $S \subseteq S'$. We modify all
program instructions to be extensive using the following rules.
\newpage
\begin{itemize}

  \item An assignment $\mathtt{x = y}$ is transformed into a
  non-deterministic choice between the old and the new value as $\mathtt{x
  = (\ast) \;?\; x : y}$.

  \item A function call $\mathtt{f(x_1, \ldots, x_n)}$ where at least one of the
  arguments $x_i$ is passed by reference is transformed into a condition which
  non-deterministically skips the call as $\mathtt{if(\ast)\; f(x_1, \ldots,
  x_n)}$.

\end{itemize}

%\footnotetext{\scriptsize{{Inspired by
%\url{https://gitlab.com/sosy-lab/benchmarking/sv-benchmarks/-/blob/d068d261f99b153c601e4d235eecec5f11ee012e/c/pthread/singleton.c}.}}}

%\footnotetext{\scriptsize{{Inspired by
%\url{https://gitlab.com/sosy-lab/benchmarking/sv-benchmarks/-/blob/b7118e858cb0f0648536a40bef545941d4ac3a25/c/goblint-regression/13-privatized_03-priv_inv.c}}}}

In the
rest of the paper, we implicitly assume that all backend queries for a thread
$t$ are evaluated over the results $R_i(t)$ where $R_i$ is the value returned by
Algorithm \ref{alg:thread-analysis}.

%%%%%%%%%%%%%%%%%%%%%%%%%%%%%%%%%%%%%%%%%%%%%%%%%%%%%%%%%%%%%%%%%%%%%%%%%%%%%%%%
\section{Lockset and Active-Threads Analyses}
\label{sec:lockset_and_thread_analyses}
%%%%%%%%%%%%%%%%%%%%%%%%%%%%%%%%%%%%%%%%%%%%%%%%%%%%%%%%%%%%%%%%%%%%%%%%%%%%%%%%

This section presents two helper analyses for data race detection. Both are used
to check whether a pair of contexts can happen in parallel -- the
\textit{lockset analysis} checks this based on locks being held in the given
contexts, and the \textit{active-thread analysis} deals with creating and
joining of threads.

%===============================================================================
\subsection{Lockset Analysis}
\label{sec:lockset_analysis}
%===============================================================================

Lockset analysis is traditionally used to compute must- or may-locksets being
held for each statement. Our lockset analysis is  context-sensitive,
path-insensitive, and it computes both must- and may-locksets simultaneously. It is
inspired by the approach of \racerX described in \cite{racerx}, which, however,
does not employ any advanced alias analysis and is not context-sensitive.

We represent locks by their concrete addresses, i.e., $\mathsf{Lock} =
\addresses$, and we define $\locksets = 2^\mathsf{Lock}$. Our lockset analysis
is implemented as a forward data flow analysis with the domain of locksets,
computed for each thread in separation, starting with the empty lockset. On the
intraprocedural level, we start to analyse a function with an entry lockset
$L_{entry}$ coming from the caller, and gradually update it using the following
transfer function that, for a context $c$ with a statement $s$,
%and using $P =\mathsf{Backend.value\_ptr}(c, p)$
transforms the lockset $L_{old}$ reached
before executing $s$ into a set of locksets $\mathcal{L}_{new}$ resulting from
the execution of $s$ as follows:

%$\mathcal{L}_{new} = L_{old} \cup \{\ell\}
%\;|\; \ell \in P\}$ if $s$ is $\mathtt{lock}(p)$, $\mathcal{L}_{new} =  \{
%L_{old} \setminus \,\{\ell\} \;|\; \ell \in P\}$ if $s$ is $\mathtt{unlock}(p)$,
%and $\mathcal{L}_{new}  =\{L_{old}\}$ otherwise.

\begin{align*}
    \mathcal{L}_{new} =
        \begin{cases}
        \{ L_{old} \cup \{\ell\} \;|\; \ell \in \mathsf{Backend.value\_ptr}(c, \mathtt{p})\} &
          \textrm{if s is $\mathtt{lock(p)}$},\\
        %\{ L_{old} \setminus \,\{\ell\} \;|\; \ell \in P\} \cup
        % \{\emptyset\} & \textrm{if s is \texttt{unlock}(p) and $|P| > 1$,}\\
        \{ L_{old} \setminus \,\{\ell\} \;|\; \ell \in \mathsf{Backend.value\_ptr}(c, \mathtt{p})\} & \textrm{if s
          is $\mathtt{unlock(p)}$,}\\
        \{L_{old}\} & \textrm{otherwise.}
        \end{cases}
\end{align*}

% \noindent Notice that in the second case, when we cannot precisely determine a
% lock to be unlocked, we also consider conservatively consider a case that
% every lock was unlocked.

After applying the transfer function, if $\mathcal{L}_{\mathrm{new}}$ is not a
singleton set, we fork the analysis and continue with each element of
$\mathcal{L}_{\mathrm{new}}$ separately. As a result, we obtain the summary of a
function $f$ analysed in a context $c$ in the form $(c, L_{entry} )\mapsto
\mathcal{L}_{exit}$ where $\mathcal{L}_{exit}$ is the set of locksets at the
return statements of all forked analysis runs inside~$f$. Such summaries are
then used for the interprocedural analysis. We also construct \emph{statement
summaries} $\mathcal{S}$ of the same form as for functions, i.e., mappings
from pairs of an input context and a lockset to output sets of locksets. These
are used for caching at the statement level and to implement the must- and
may-lockset queries for a context $c$ as

%$\mathsf{may\_ls}(c) = \bigcup_{(c, L)
%\in \mathsf{dom}(\mathcal{S})} L$ and $\mathsf{must\_ls}(c) = \bigcap_{(c, L)
%\in \mathsf{dom}(\mathcal{S})} L$.

\begin{align*}
    \mathsf{may\_ls}(c) = \bigcup_{(c, L) \in \mathsf{dom}(\mathcal{S})} L && \mathsf{must\_ls}(c) = \bigcap_{(c, L) \in \mathsf{dom}(\mathcal{S})} L
\end{align*}

%-------------------------------------------------------------------------------
\subparagraph*{Dealing with path explosion.}
%-------------------------------------------------------------------------------

As we represent locks using memory addresses and fork the analysis whenever we
cannot precisely evaluate the parameter of a (un)locking operation, the set of
traversed paths can easily explode. To tackle this problem, our transfer
function selects only a small number of locks when the set
$\mathsf{Backend.value\_ptr}(c, \mathtt{p})$ is too large (currently, the threshold is set
to three locks). This works in many practical cases because it is usually enough
to differentiate the may- and must-cases through a few paths only. In the
future, we would like to provide support for symbolic treatment of locks
combined with may- and must-equality reasoning.\\

Our analysis does not support reentrant mutexes/semaphores but can handle
non-blocking lock functions and read-write locks as discussed below.

%-------------------------------------------------------------------------------
\subparagraph*{Non-blocking locking functions.}
%-------------------------------------------------------------------------------

For non-blocking lock functions such as \texttt{trylock}, which does not block
when the mutex is already locked, or \texttt{timedlock}, which blocks just for a
limited time, we update the lockset with relevant locks for which we also track
the variable used to store the return value of the corresponding call.
Evaluation of a query $\mathsf{may\_ls}(c)$ or $\mathsf{must\_ls}(c)$ will then
ask the backend for the state $s$ computed for the context $c$ through the query
$s = \mathsf{Backend.state}(c)$, and filter out the locks for which the state
$s$ does not guarantee that the locking was (possibly or surely, depending on
the query) successful. This guarantee will only be provided if the success of
the locking was explicitly checked in the context $c$. This way we obtain a
limited path-sensitivity for branching over expressions containing variables
storing return values of lock operations.  For the function
\texttt{pthread\_mutex\_lock}, we assume that it is blocking (i.e., that it
never fails), but the user can easily change the configuration to treat it as
non-blocking.

%-------------------------------------------------------------------------------
\subparagraph*{Read-write locks.}
%-------------------------------------------------------------------------------

To support read-write locks, a lock object keeps additional information about
whether it was obtained in the read- or write-mode. Based on this information,
we modify the intersection of locksets, presented later on as the main operation
used for data race detection, so that it contains only those locks which are
present in both operands and at least in one of them in the write-mode. In other
words, if two accesses are guarded only by a lock obtained in both cases in the
read-mode, we will treat them as unprotected.

%===============================================================================
\subsection{Active-Threads Analysis}
%===============================================================================

The goal of this analysis is to deal with situations when two memory accesses
cannot happen in parallel either because one happens before the thread of the
other one is created, or because one happens after the thread of the other one
is surely joined. We again compute this information in a thread-modular way by
computing a \emph{set of sets of threads} that may or must be active in  parallel for
each context, respectively. Two contexts are then may-parallel (must-parallel)
if the thread of one of them is in the union (intersection) of sets computed for
the other, and vice versa.

The sets of active threads are computed in a very similar way to how the lockset
analysis is performed -- just working with sets of threads instead of sets of
locks, with the transfer functions for thread creation and joining analogous to those for locking/unlocking, and with the same way of handling the may- and
must-queries. The set of thread entry points is usually small, and we thus do
not need any special treatment to tackle path explosion unlike in the lockset
analysis. One difference between the analyses is in the initial states. The
lockset analysis of each thread is started with the empty lockset, but the
active-threads analysis of a thread $t$ starts with the union of all threads
that may be active together with $t$ in some context of the parent of $t$ (the
analysis is thus run in the order given by a topological order on
thread-creation relation, and for the very rare case of cyclic thread-creation
graphs, we simply skip the analysis, providing trivial answers for queries based
on it).

\subparagraph*{Thread identifiers.}

Another significant difference is the way in which threads are joined. While our
analysis abstracts threads using their entry points, in the concrete semantics,
they are manipulated using their \textit{identifiers}. In many programs,
however, the manipulation of identifiers is straightforward, and one can easily
map identifiers to abstract threads and vice versa. Based on this observation,
we assume that if an identifier expression of a created and joined thread is the
same, then the thread is also the same. We have never encountered a situation
when this would lead to a wrong analysis verdict.

%%%%%%%%%%%%%%%%%%%%%%%%%%%%%%%%%%%%%%%%%%%%%%%%%%%%%%%%%%%%%%%%%%%%%%%%%%%%%%%%
\section{Data Race Detection}
\label{sec:race-detection}
%%%%%%%%%%%%%%%%%%%%%%%%%%%%%%%%%%%%%%%%%%%%%%%%%%%%%%%%%%%%%%%%%%%%%%%%%%%%%%%%

Our data race detection does a forward traversal of the program and collects
memory accesses consisting of a context (used for querying the results of
lockset and active-thread analyses), a base address, and a symbolic offset (a
set of intervals in the \textsf{Cvalue} domain that may be some simple value for
backends different from EVA). We cluster memory accesses according to their
bases to later check for data races only inside each of those clusters. This is
justified by the following hypothesis about the memory model: it is not possible
to get from one base to another by adding an offset. This hypothesis might not
hold for some low-level C programs, but it is important for efficiency.
Moreover, analysis of programs breaking this assumption is not even supported by
EVA \cite{eva-user-manual}.

The tracking of accesses is inspired by \cite{Eraser}. For each cluster
represented by its base, we track a state $s \in \{\RO\;o, \EX\xspace\;o, \SH,
\SM\}$ representing that a memory base is either read-only (\RO) or modified
(\EX) by a single thread $o$, read by several threads with none of them writing
to it (\SH), or modified by distinct threads such that at least one of them
writes to it (\SM). With each access to a memory base $b$ by a thread $t$, the
state of $b$ is updated according to the transition graph in Fig.
\ref{fig:state_transitions}. When the analysis is finished, we keep only the
memory bases marked as \SM\xspace or \mbox{\EX\;$o$ } (when $o$ is a thread
created multiple times), and for each of them, we check each pair of accesses
for a data race. The information about which threads are unique and which are
created multiple times is computed during the thread analysis based on whether
its create statement was visited once or multiple times.

\begin{figure}
     \begin{tikzpicture}[shorten >=1pt,node distance=3cm,auto]
        \tikzstyle{state}=[thick,draw=black,fill=green!10, minimum size=10mm,rounded corners=0.15cm]

        \node[state, align=center] (R) [xshift=-10mm] 
            {$\mathsf{ReadOnly}$\\(\textit{owner})}; 
            
        \node[state, align=center, fill=red!10] (E) [above of=R, yshift=5mm] 
            {$\mathsf{Exclusive}$\\(\textit{owner})}; 
            
        \node[state] (S) [right of=R, xshift=25mm] 
            {$\mathsf{Shared}$}; 
            
        \node[state, fill=red!10] (SM) [above of=S, yshift=5mm] 
            {$\mathsf{SharedMod}$};

        \node[left of=R, xshift=0cm] (I1) {$\;$};
        
        \node[left of=E, xshift=0cm] (I2) {$\;$};

        %% Edges

        \draw[-stealth'] (I1) -- node [below, align=center, xshift=-5mm] {$\mathsf{read}$, t = owner} (R);

        \draw[-stealth'] (I2) -- node [above, align=center, xshift=-5mm] {$\mathsf{write}$, t = owner} (E);
    
        \draw[-stealth'] (R) to [loop below] (R) node [below=33pt] 
            {$\mathsf{read}$, t = owner};

        \draw[-stealth'] (E) to [loop above] (E) node [above=33pt] 
            {$\ast$, t = owner};

        \draw[-stealth'] (SM) to [loop above] (SM) node [above=33pt] 
            {$\ast$};

        \draw[-stealth'] (S) to [loop below] (S) node [below=33pt] 
            {$\mathsf{read}$, $\ast$};

        \draw[-stealth'] (R) -- node [below, align=center] {$\mathsf{read}$, t $\neq$ owner} (S);
        \draw[-stealth'] (R) -- node [left, align=center] {$\mathsf{write}$, t $=$ owner} (E);
        \draw[-stealth'] (E) -- node [above, align=center] {$\ast$, t $\neq$ owner} (SM);
        \draw[-stealth'] (S) -- node [right, align=center] {$\mathsf{write}$, $\ast$} (SM);

        \draw[-stealth'] (R.80) -- node [right, xshift=-3mm, yshift=-3mm, align=center] {$\mathsf{write}$,  t $\neq$ owner} (SM.190);

\end{tikzpicture}
    \centering
    \caption{The transition graph for tracking states of memory bases. Each
    transition corresponds to a \textsf{read} or \textsf{write} memory access by
    a thread $t$. The \mbox{symbol $\ast$} represents any access kind/thread.
    The states vulnerable to data races are marked in red.}
    \label{fig:state_transitions}
\end{figure}

%-------------------------------------------------------------------------------
\subparagraph*{Escape analysis.}
%-------------------------------------------------------------------------------

An important part of data race detection is discovering local bases that may
escape their scope. Our syntactic backend assumes that dynamically allocated
bases can always escape and static bases escape when their address
is taken. Using the Alias and EVA backends, we enhance this by testing (for both static and dynamic bases) not only whether their addresses were taken, but also whether they were assigned to global variables.

%-------------------------------------------------------------------------------
\subparagraph*{May- and must-data-races.}
%-------------------------------------------------------------------------------

We distinguish two types of data races based on the confidence we have in them.
For \textit{may-data-races}, the below conditions need to be
satisfied:

\begin{enumerate}
  \item At least one of the accesses is a write access.
  \item The accessing threads are either distinct or the same non-unique thread.
  \item The offsets of the addresses may overlap (checked by a query over the
  \textsf{CValue} domain of intervals).
  \item The intersection of the may-locksets of the concerned contexts is empty
  (checked using the results of the lockset analysis).
  \item The threads of both accesses may run in parallel (checked using the
  results of the active-threads analysis).
\end{enumerate}

For \textit{must-data-races}, we require the \textit{must}-variants of
Conditions 3--5 to hold. In addition, the following conditions are required to
rule out races involving a memory access representing abstraction of multiple
memory accesses:

\begin{enumerate}
  \setcounter{enumi}{5}
  \item The memory base is not \textit{weak} (i.e., if it is dynamic, then
  it represents precisely one dynamic allocation) and it is not allocated in a
  non-unique thread.
  \item The size of the access does not exceed the size of the type of the
  base.
  \item The base is the only base being read/written to in its context.
\end{enumerate}

\noindent Condition~6 ensures that we do not report must-races on bases obtained
by abstracting and hence merging different calls to allocation functions.
Condition~7 rules out races on possibly different indices to arrays (an access
with an offset bigger than the size of the type of the base usually abstracts
multiple accesses). Finally, Condition~8 handles situations when our
context-sensitivity level is not sufficient to distinguish accesses to multiple
bases.

%-------------------------------------------------------------------------------
\subparagraph*{Arrays and loop unfolding.}
%-------------------------------------------------------------------------------

To slightly relax the restrictiveness of Condition~7,  we can perform a small
number of unfoldings of each loop. This can help us report must data races on
the accesses in unfolded cases. The loop unfolding is also useful for dealing
with the creation of threads in loops -- since our analyses are
path-insensitive, they will also consider the infeasible path on which the cycle
is never entered, and no thread is created. For memory accesses following the
loop, the must variant of Condition~5 will thus not hold.  Unfolding such loops
at least once will fix this problem.

%%%%%%%%%%%%%%%%%%%%%%%%%%%%%%%%%%%%%%%%%%%%%%%%%%%%%%%%%%%%%%%%%%%%%%%%%%%%%%%%
\section{Implementation and Experimental Evaluation}
\label{sec:experiments}
%%%%%%%%%%%%%%%%%%%%%%%%%%%%%%%%%%%%%%%%%%%%%%%%%%%%%%%%%%%%%%%%%%%%%%%%%%%%%%%%

The \racerF\footnote{\racerF is available under the MIT license at
\url{https://github.com/TDacik/Deadlock_and_Racer}.} 
analyser based on the above
presented principles is implemented as a plugin of the Frama-C platform as a
part of our bigger project for detection of concurrency bugs. The individual
analyses are written as OCaml functors parametrised by a module implementing the
backend analysis. We also provide a Python wrapper which runs a combination of
under- and over-approximating analyses and also performs preprocessing
(currently, the preprocessing is implemented outside of our analyser for
technical reasons, thus the analyser is slowed a bit by the need of repeated
parsing). To solve the systems of equations from Sections
\ref{sec:thread-analysis_under} and \ref{sec:thread-analysis_over} we use the
\emph{chaotic iteration strategy with widenings}
\cite{Bourdoncle1993EfficientCI} implemented in the Ocamlgraph library
\cite{ocamlgraph}. The widening operators are those provided by the
\textsf{CValue} domain.

Our implementation provides support for atomic and thread-local variables. The
threading and locking functions (and their relevant properties) as well as
atomic functions (such as gcc's  \texttt{\_\_atomic\_store}) may be defined
using configuration files.

We have conducted a series of experiments to support the following two
hypotheses:\begin{itemize}

  \item Despite having no formal guarantees, \racerF can efficiently analyse
  intricate programs without sacrificing a precision. We validated this
  hypothesis on a collection of benchmarks from SV-COMP containing 1029
  programs, many of them being intricate examples and corner cases contributed
  by participants to showcase capabilities of their tools.

  \item \racerF can detect data races in real-world programs for which other
  existing static analysers do not provide precise results (either because of too coarse
  approximation or running out of resources). We  validated this hypothesis on a
  set of student programs from \cite{anaconda}.

\end{itemize}

%-------------------------------------------------------------------------------
\subparagraph*{Experimental setup.}
%-------------------------------------------------------------------------------

Some data presented in experiments were taken from the results of SV-COMP 2025
\cite{SVCOMP25}. All other experiments were run on a machine with 2.5 GHz Intel
Core i5-7300HQ CPU and 16 GiB RAM, running Ubuntu 22.04 in a virtual machine
limited to 10.7 GiB RAM. Our experiments were conducted using
BenchExec~\cite{benchexec}, a framework for reliable benchmarking.

%===============================================================================
\subsection{SV-COMP Benchmarks}
\enlargethispage{6mm}
%===============================================================================

The SV-COMP sub-category \textit{NoDataRace-Main} is, to our knowledge,
the largest publicly available data race benchmark. Many of its programs are
small but intricate tests provided by the participants. Our main goal is to
demonstrate the scalability of our approach and to also show that its heuristic
nature does not compromise its precision even on those intricate examples. The
benchmark also contains implementations of lock-free algorithms, which are out
of the scope of our tool. We have implemented a single heuristic that detects
active waiting (essentially a loop with an empty body) as a characteristic
pattern for lock-free algorithms, and when it is present, returns an unknown
result.

We compare with the five best-performing tools participating in SV-COMP 2025
\cite{deagle,UAUTOMIZER-SVCOMP24,gemcutter,GOBLINT-SVCOMP24} (including \racerFsv, an older version
of \racerF) and we also include \racerFunder and \racerFover representing
\racerF running with under- and over-approximating thread analysis only, respectively. We
wanted to include also the \otwo checker~\cite{o2}, but since its results were
not good and all the other tools were configured specifically for this benchmark
by their authors, we do not consider such a comparison fair.

\begin{table*}[!t]
    \centering

    \caption{Results for the SV-COMP sub-category \textit{NoDataRace-Main}. The columns
    represent true negatives, false positives, true positives, false negatives,
    out-of-resource (time or memory), score (as defined by SV-COMP rules, but excluding witness validation), and times (excluding timeouts) on local and competition
    machine, respectively.}

    \label{table:svcomp}
    \begin{tabular}{@{}lrrcrrrrrrr@{}}
\toprule
& \multicolumn{2}{c}{No race (794)} & & \multicolumn{2}{c}{Races (235)}\\

\cmidrule{2-3} \cmidrule{5-6}
Analyser$\hspace{30pt}$ & TN & FP & $\;$ & TP & FN & RO & Score &Time$_1$ [s]& Time$_2$ [s]\\
\midrule
\rowcolor{GreenYellow}

\textsc{RacerF} & 674 & 4 & & 105 & 0 & 0 & 1389 & 4620 & -\\
\textsc{RacerF}$_\mathrm{over}$ & 674 & 6 & & 83 & 0 & 0 & 1335 & 3970 & -\\
\rowcolor[gray]{0.9}
\textsc{RacerF}$_\mathrm{under}$ & 680 & 4 & & 105 & 6 & 0 & 1209 & 2710 & -\\
\midrule
\textsc{RacerF}$_\mathrm{sv-comp}$ & 674 & 4 & & 98 & 0 & 0 & 1382 & 3100 & 1300\\
\rowcolor[gray]{0.9}
\textsc{Goblint} & 712 & 0 & & 0 & 0 & 3 & 1424 & - & 2320\\
\textsc{Deagle} & 685 & 0 & & 203 & 1 & 43 & 1541 & - & 10k\\
\rowcolor[gray]{0.9}
\textsc{UGemCutter} & 578 & 0 & & 161 & 0 & 263 & 1317 & - & 152k\\
\textsc{UAutomizer} & 609 & 0 & & 121 & 0 & 274 & 1339 & - & 120k\\
\bottomrule
\end{tabular}
\end{table*}

%\begin{wrapfigure}{r}{0.5\textwidth}
%    \includegraphics[width=0.5\textwidth]{figs/kaktus.png}
%    \caption{\td{Placeholder. Use a prettier graph if we have time to create it.}}
%    \label{fig:cactus}
%    \vspace{-15pt}
%\end{wrapfigure}

The results are summarised in Table~\ref{table:svcomp}. Unlike the competition
results \cite{SVCOMP25}, we present the results ignoring the witness validation.
\racerF produced 4 false alarms, all of them on programs where a data race is
ruled out by some intricate condition. Two of them are variants of a program
where a data race does not manifest because an array being searched in
parallel contains only zeros and a certain condition is thus never triggered which makes the potential racy access unreachable
(which is the kind of behaviour that our tool cannot handle). The other two contain locking patterns that are beyond the scope of our lockset analysis. 

We first focus on a comparison of the versions of \racerF. The difference
between its newest version and the version competing in SV-COMP'25 lies mainly
in improved escape analysis and preprocessing (loop unrolling in the
under-approximating mode and assignment modification in the over-approximating
mode). The changes show a positive impact on more reported data races but also
a slowdown which is a consequence of preprocessing done in over-approximating
mode. The slowdown mostly manifests for the active-thread analysis, which could
be a performance bug in the implementation.

Interestingly, running \racerF in either the over- or under-approximating mode
does not compromise its precision too much, leading to an additional 2 false
positives (and 22 must-races classified just as may- and thus not reported) and
6 false negatives, respectively.
%However, we are aware that there are more
%programs where approximation could make a difference, but \racerF returns an
%unknown result for another reason.

Compared to participants of SV-COMP'25, \racerF placed second in the competition
ranking and third in our evaluation. Compared to \goblint, which never reports
races, \racerF can produce correct results for more programs. Compared to model checker
\deagle, which benefits from the fact that many programs are small and their
loops can be fully unrolled, no version of \racerF ever runs out of time or memory.

Where \racerF shines is the
\textsf{ldv\!\!-\!\!linux\!\!-\!\!3.14\!\!-\!\!races} benchmark subset
containing 5~programs derived from the Linux kernel (all of them are
race-free). None of the participants with a positive score can provide
correct verdicts for those programs (except \goblint which handles one of them),
but \racerF can correctly analyse each of them under two minutes.

\begin{table*}[!t]
    \centering
    \caption{Results for student programs. The columns are true positives, false negatives, no race reported, race reported, out-of-resources, and time (excluding timeouts).}
    \label{table:pos}
    \begin{tabular}{@{}lrrcrrrr@{}}
\toprule
& \multicolumn{2}{c}{Races (30)} & & \multicolumn{2}{c}{Others (86)}\\

\cmidrule{2-3} \cmidrule{5-6}
Analyser$\hspace{15pt}$ & $\;$TP$\;$ & $\;$FN & $\;\;\;\;$ & $\;$NR$\;$ & $\;\;\;\;$RR$\;$ & $\;\;\;$OOR & $\;\;\;$Time [s]\\
\midrule
\rowcolor{GreenYellow}
\textsc{RacerF} & 16 & 3 & & 46 & 0 & 0 & 688 \\
\textsc{O2} & 7 & 23 & & 86 & 0 & 0 & 54 \\
\rowcolor[gray]{0.9}
\textsc{Goblint}$_\mathrm{sv}$ & 0 & 3 & & 46 & 0 & 0 & 41 \\
\textsc{Goblint} & 27 & 3 & & 46 & 40 & 0 & 41 \\
\rowcolor[gray]{0.9}
\textsc{Deagle} & 3 & 1 & & 6 & 0 & 78 & 1644\\

%\midrule
%\rowcolor{GreenYellow}
%\textsc{RacerF} & 674 & 4 & & 110 & 0 & 0 & 1394 & 0 & 4480 & -\\
%\textsc{RacerF}$_\mathrm{over}$ & 674 & 6 & & 87 & 0 & 0 & 1339 & 0 & 3810 & -%\\
%\rowcolor[gray]{0.9}
%\textsc{RacerF}$_\mathrm{under}$ & 680 & 4 & & 110 & 6 & 0 & 1208 & 0 & 2630 & -\\
%\midrule
%\textsc{RacerF}$_\mathrm{sv-comp}$ & 674 & 4 & & 98 & 0 & 0 & 1382 & 0 & 3100 & 1300\\
%\rowcolor[gray]{0.9}
%\textsc{Goblint} & 712 & 0 & & 0 & 0 & 3 & 1424 & 0 & - & 5200\\
%\textsc{Deagle} & 685 & 0 & & 203 & 1 & 43 & 1541 & 0 & - & 44k\\
%\rowcolor[gray]{0.9}
%\textsc{GemCutter} & 578 & 0 & & 161 & 0 & 263 & 1317 & 0 & - & 290k\\
%\textsc{Automizer} & 609 & 0 & & 121 & 0 & 274 & 1339 & 0 & - & 170k\\
\bottomrule
\end{tabular}

\end{table*}

%===============================================================================
\subsection{Student Programs}
\enlargethispage{6mm}
%===============================================================================

To test our tool on more non-crafted programs, we performed an evaluation on a
set of 116 student programs from \cite{anaconda}. The programs are rather short
(around 100 -- 300 LoC) and implement the same heavily concurrent algorithm
parametric in the number of threads, but they differ in various intricate bugs
(for their detailed discussion see \cite{anaconda}). We have identified 29 data
races using dynamic analysers. However, there may be more that those tools
missed. One additional race was discovered by all \racerF, \otwo, and \goblint. The
issue could manifest only in a very rare situation when two threads
simultaneously failed and called a clean-up function that assigns a global
pointer without synchronization. The time limit was set to 60 seconds and the
memory limit to 1GB.

We have compared with \goblint\footnote{We use \goblint in the version from
SV-COMP'25 but with the default configuration as the other available
configurations did not improve the results and led to timeouts. Further, to
provide more detailed results, we run \goblint via its SV-COMP wrapper that
never reports detected races \mbox{(\goblint$_{\!\mathrm{sv}}$)} as well as without
it.} and \deagle (version from SV-COMP'25) that performed better than us in
SV-COMP, and with O2 (implemented in Coderrect\footnote{The link
\url{https://coderrect.com} from which we got version 1.1.3 is no longer
available.}) that we selected as a recent representative of lightweight and
scalable race detection tools.

The results are given in Table \ref{table:pos}. As expected, since all the programs
contain unbounded loops and also an unbounded number of threads is created, \deagle ran
out of the resources on almost all programs and provided an answer in only 10 cases.
Interestingly, \goblint and \racerF claimed \textit{the same }46 programs as race-free (but \racerF without providing any false positives).
The three missed
races by \racerF and \goblint are also the same and are caused by an undefined behaviour related to re-initialisation and destroying of mutexes that still may be used by some threads (neither of the tools supports detection of such problems).
%with lock initialisation that the tools do not support.
A~detailed inspection
suggests that both tools struggle with programs that use an array where each
thread accesses only the element at its dedicated index.

Overall, \racerF achieves excellent precision on both sides. While O2 misses
many races and \goblint reports many false alarms, \racerF is able to discover
more than half of the races while not reporting any false alarms.

%%%%%%%%%%%%%%%%%%%%%%%%%%%%%%%%%%%%%%%%%%%%%%%%%%%%%%%%%%%%%%%%%%%%%%%%%%%%%%%%
\section{Conclusions and Future Work}
\enlargethispage{4mm}
%%%%%%%%%%%%%%%%%%%%%%%%%%%%%%%%%%%%%%%%%%%%%%%%%%%%%%%%%%%%%%%%%%%%%%%%%%%%%%%

We have presented a novel data race detector \racerF and performed a
series of experiments indicating that \racerF achieves a nice compromise
between precision and scalability both on crafted benchmarks and real-world
programs. In the future, we would like to target other classes of concurrency
bugs. We already have an implementation for deadlock detection based on top of
our lockset analysis, which is not discussed in this paper. Another interesting
direction is combination with dynamic approaches, e.g., guiding insertion of
noise injection by results of static analysis.

\bibliography{main}

\end{document}